\documentclass[preprint,12p]{elsarticle}

\usepackage{graphicx}
\usepackage{dcolumn}
\usepackage{bm}
\usepackage{epstopdf}
\usepackage{calrsfs}
\usepackage{amsmath}
\usepackage{amssymb}
\usepackage{placeins}
\usepackage{xcolor}

\begin{document}

\begin{frontmatter}

\title{The effect of time delay for synchronisation suppression in neuronal
networks}

\author{Matheus Hansen$^1$, Paulo R. Protachevicz$^2$, Kelly C. Iarosz$^{3,4}$, 
Iber\^e L. Caldas$^2$, Antonio M. Batista$^{5,6}$ and Elbert E. N. Macau$^1$}
\address{$^1$Institute of Science and Technology, Federal University of
S\~ao Paulo - UNIFESP, S\~ao Jos\'e dos Campos, S\~ao Paulo, SP, Brazil.\\
$^2$Institute of Physics, University of S\~ao Paulo, S\~ao Paulo, SP, Brazil.\\
$^3$Faculdade de Tel\^emaco Borba, FATEB, Tel\^emaco Borba, Paran\'a, Brazil.\\
$^4$Graduate Program in Chemical Engineering Federal Technological University of
Paran\'a, Ponta Grossa, PR, Brazil\\
$^5$Post-Graduation in Science, State University of Ponta Grossa, Ponta Grossa,
PR, Brazil.\\
$^6$Mathematics and Statistics Department, State University of Ponta Grossa,
Ponta Grossa, PR, Brazil.}
\cortext[cor]{mathehansen@gmail.com}

\date{\today}

\begin{abstract}
We study the time delay in the synaptic conductance for suppression of spike
synchronisation in a random network of Hodgkin Huxley neurons coupled by means
of chemical synapses. In the first part, we examine in detail how the time
delay acts over the network during the synchronised and desynchronised neuronal
activities. We observe a relation between the neuronal dynamics and the syaptic
conductance distributions. We find parameter values in which the time delay has
high effectiveness in promoting the suppression of spike synchronisation. In the
second part, we analyse how the delayed neuronal networks react when pulsed
inputs with different profiles (periodic, random, and mixed) are applied on the
neurons. We show the main parameters responsible for inducing or not synchronous
neuronal oscillations in delayed networks.
\end{abstract}

\end{frontmatter}


\section{Introduction}
\label{sec1}

The understanding of emergence, as well as the control, of neuronal
synchronisation is one of the central points of contemporary neuroscience,
mostly due to the fact that synchronous patterns can be related to some
fundamental neuronal processes for life, such as memory \cite{gruber18}, 
perception \cite{ jamal15} and also to some brain disorders, for instance
epilepsy \cite{traub82}. Specifically considering the case of brain pathologies,
many studies have been developed in order to find alternative methods to control
synchronous neuronal activities. Protachevicz et al. \cite{paulo19} indicated 
through numerical simulations that external perturbations can not only induce
peak synchronisation, but also reduce the abnormal synchronous pattern.
Recently, Cota et al. \cite{cota21} observed in experimental analyses with rats
that nonperiodic electrical stimulation can be a very promising alternative for
the treatment of epileptic seizures. Inspired by this scenario, this work aims 
to contribute to the exploration of the effects of delayed conductance on
neuronal synchronisation activities. We focus on cases in which the temporal
delay has a positive performance in suppressing spike synchronisation.

In neuronal communication, the time delay is an intrinsic property, being
associated with axonal, dendritic, and synaptic signal propagation
\cite{asl2017,asl2018}. In the axons, the presence of myelination is responsible
for the rapid signal transmission \cite{seidl2014}, while the demyelination
causes a reduction in the conduction velocity \cite{waxman1977}. In the chemical
synapses, the time delay is between less than one millisecond
and up to tens of milliseconds \cite{purves2018,stoelzel2017}. For the
dendritic, the time delay is smaller than one millisecond \cite{spencer2012}. 

As a model to mimic neuronal activities, we use the Hodgkin Huxley (HH)
\cite{hodgkin52} neuron, that was proposed in $1952$ by physiologists Alan
Hodgkin and Andrew Huxley. This model was developed in a successful attempt to
describe the mechanisms of action potential generation in experiments with 
the giant squid axon. In such a model it was reported that the generation of the
action potential in the cell membrane is linked to variations in the ionic
currents of potassium, sodium and a current defined by them as leak. Although
varous mathematical models have been proposed to reproduce the neuronal 
behaviour \cite{rulkov,idex}, the HH neuron is still one of the most actual
approaches to neuronal dynamics,  inspiring several studies in the field of
neuroscience \cite{popovych14,borges17}.

Our main finding in this work is to show the mechanism responsible for the
emergence of spike synchronisation in networks composed of HH neurons, randomly
coupled by means of chemical synapses. We explore the effects of delayed
conductance on neuronal activities as an alternative method for suppressing or 
reducing of synchronous patterns. Such analyses are carried out in two
different scenarios, where firstly the neuronal network has no external
perturbation on the inputs, and secondly when pulse perturbations (for instance
sensory sensory stimulation) with periodic, random, and mixed profiles are 
considered \cite{hansen22}. In both cases, we discuss the conditions and
parameters in which the time delay is able or not to hold low levels of spike
synchronisation in neuronal networks.

The paper is organised as follows. In Section 2, we introduce the mathematical
model of coupled HH neurons. In Section 3, we present the main diagnostics used 
in the study. Section 4, we exhibit the suppression of synchronised activities 
under constant current input due to the time delay. Section 5 shows the time 
delay effect under pulsed perturbed neuronal networks. Finally, we highlight
the conclusions of our work in the last section.


\section{Neuron model}
\label{sec2}

The HH model \cite{lameu21} with time delay is given by
\begin{eqnarray}
C\frac{dV_i}{dt} & = & -g_{\rm K}n_i^4(V_i-V_{\rm K})-g_{\rm Na}m_i^3h_i(V_i-V_{\rm Na})
\nonumber \\  
& & -g_{\rm l}(V_i-V_{\rm l})+I_{i}+I_i^{\rm syn}(t-\tau), \\
\frac{dx_i}{dt} & = &\alpha_{x_i}(v_i)(1-x_i)-\beta_{x_i}(v_i)x_i,
\end{eqnarray}
where $C$ is the capacitance of the cell membrane, $V_i$ is the membrane
potential for the $i$-th HH neuron, and $t$ is the time. The parameters
$g_{\rm K}$, $g_{\rm Na}$, and $g_{\rm l}$ correspond to the potassium, sodium, and
leak maximal conductance, respectively. The variables $n_i$ and $m_i$ are
related to the possibility of the ionic channels of potassium (K$^+$) and
sodium (Na$^+$) be open (active), while $h_i$ is associated with the possibility
of the sodium channel (Na$^+$) be close (inactive). $V_{\rm K}$, $V_{\rm Na}$, and
$V_{\rm l}$ represent the potassium, sodium, and leak reversal potential, while
$I_i$ and $I_{i}^{\rm syn}$ correspond to the external and synaptic current
density, respectively. In order to simplify the mathematical expression of
opening and closing channels, we condense them as represented in Eq. (2), where
$x_i$ can be $n_i$ , $m_i$, and $h_i$. $\alpha_{xi}$ and $\beta_{xi}$ are
different functions of $v_{i}$ that depend on $n_{i}$, $m_{i}$, and $h_{i}$. In
this equation, $v_{i}=V_{i}/[mV]$ represents the value of the dimensionless
membrane potential. The $\alpha_{xi}$ and $\beta_{xi}$ are experimental functions
found by Hodgkin and Huxley and written as
\begin{eqnarray}
\alpha_n(v_i) & = & \frac{0.01v_i+0.55}{1-{\rm exp}(-0.1v_i-5.5)}, \\ 
\alpha_m(v_i) & = & \frac{0.1v_i+4}{1-{\rm exp}(-0.1v_i-4)}, \\
\alpha_h(v_i) & = & 0.07{\rm exp}\left(\frac{-v_i-65}{20}\right), \\
\beta_n(v_i) & = & 0.125{\rm exp}\left(\frac{-v_i-65}{80}\right), \\
\beta_m(v_i) & = & 4{\rm exp}\left(\frac{-v_i-65}{18}\right), \\
\beta_h(v_i) & = & \frac{1}{1+{\rm exp}(-0.1v_i-3.5)}.
\end{eqnarray}

In this work, we consider that each $i$-th HH neuron is stimulated over time by 
an external current density $I_i=I_{i}^{\rm 0}+\xi(t)$, where $I_{i}^{\rm 0}$ is a
current density with a constant amplitude and $\xi(t)$ is a pulse with amplitude
$\Gamma$, which initially is equal to $0$ for all times (absence of pulses),
leading to $I_i=I_{i}^{\rm 0}$ (constant input). The external current density
$I_i$ is responsible for the generation of the spike dynamics. As one of our
main goals is to study spike synchronisation suppression, we uniformly
distribute $I_{i}^{\rm 0}\in[10,14]$ $\mu$A/cm$^2 $, once in this range all HH
neurons are in spike activities with inter-spike intervals (ISI) in $[13,14.6]$ 
ms \cite{hansen22}.

\begin{figure}[t]
\centerline{\includegraphics[width=0.99\linewidth]{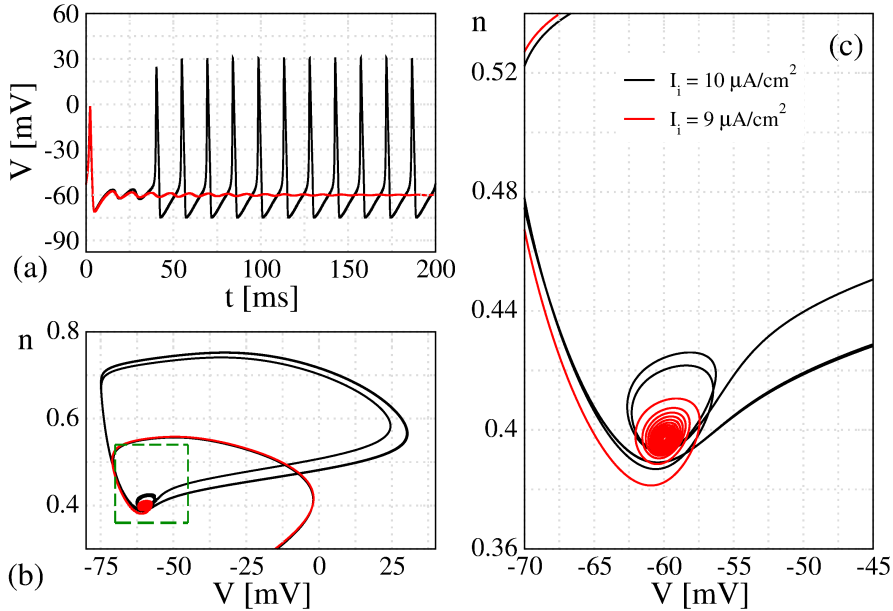}}
\caption{(a) Representation of the spike (black line) and silent (red line)
states for a single HH neuron. (b) Phase space $n\times V$ for the states
displayed in the panel (a). The black line indicates the limit cycle responsible
for the spike dynamics and the red line is the convergence to the fixed point
(silent state). (c) Magnification of the green box in the panel (b).}
\label{fig1}
\end{figure}

The behaviour of the HH neuron is separated into two different states. The first
one is the spike state that is characterised by a sudden increase in the
membrane potential value. The second state is the silent, in which the membrane
potential exhibits a small oscillation amplitude around the resting potential.
These two different patterns are displayed in Fig. \ref{fig1}(a), where the
black and red lines are the spike and silent states, respectively. The spike
dynamics can be understood when the neuron solution converges to a limit cycle
(LC), while the silent behaviour occurs due to convergence to a fixed point
(FP). The transition from one state to another is related to a Hopf bifurcation
\cite{keener98,andreev19}. Depending on the initial condition, the value of $I_i$ can be
or not enough to contribute to the bifurcation, leading the neurons from the
silent to spike states. Figure \ref{fig1}(b) displays the convergence to a limit
cycle (black line) and a fixed point (red line) in the phase space $n\times V$.
We observe that the activity of a single HH neuron $i$ depends on the external
current density $I_i$ applied over it. Figure \ref{fig1}(c) exhibits a
magnification of the green box in Fig. \ref{fig1}(b).

\begin{figure}[htbp]
\centerline{\includegraphics[width=0.5\linewidth]{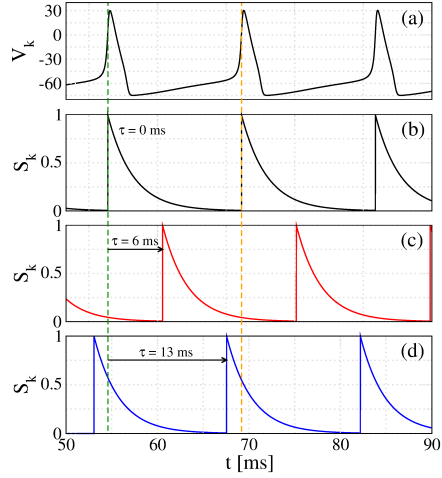}}
\caption{Schematic representation of the synaptic conductance $S_{k}$ generated
by the pre-synaptic HH neuron $k$. In the panel (a), we plot the spike dynamics
for a HH neuron $k$. The panel (b) exhibits $S_{k}$ when no time delay is
considered ($\tau=0$ ms). In the panels (c) and (d), we show the effect of time
delay $\tau=6$ ms and $\tau=13$ ms over $S_{k}$, respectively.}
\label{fig2}
\end{figure}

We build a network composed of HH neurons coupled by means of excitatory
chemical synapses. The synaptic current density received by each $i$-th HH
neuron is given by  
\begin{equation}
I_i^{\rm syn}(t-\tau)=(V_{\rm r}^{\rm exc}-V_i) \frac{g_{\rm exc}}{{N_i}}
\sum_{k=1}^{N}A_{ik}U(t-t_{k})S_{k}(t-\tau),
\label{I_syn}
\end{equation}
where $V_{\rm r}^{\rm exc}$ is the excitatory reversal potential, $g_{\rm exc}$
(mS/ cm$^2$) is the maximal excitatory synaptic conductance, $N_i$ is the number
of excitatory connections received by the neuron $i$, $N$ is the number of
neurons of the neuronal network, $A_{ik}$ is the adjacency matrix, $U(t-t_{k})$
is the Heaviside function, and $S_{k}(t-\tau)$ is a auxiliary function which
describe the temporal evolution of the synaptic conductance from the
pre-synaptic neuron $k$ to the post-synaptic neurons $i$. The $S_{k}(t-\tau)$
function is written as \cite{lameu18}
\begin{equation}
S_{k}(t-\tau)=\exp{\left[-\left({t-t_{k}-\tau}\over{\tau_{\rm s}}\right)\right]},
\label{Sk}
\end{equation}
where $t_k$ represents the times in which the the pre-synaptic neuron $k$ spikes
along the numerical simulation, $\tau$ is the time delay on the transmission of
the synaptic conductance, and $\tau_{\rm s}$ is the decay time constant on the
synaptic conductance. Essentially, $\tau$ can be related to the time needed to
the signal generated by the spike of the pre-synaptic neuron $k$ achieves the
post-synaptic neuron $i$. In our simulation for $t<t_{k}+\tau$, we assume that
$S_{k}$ is equal to $0$. Figure \ref{fig2} shows the neuronal spikes for a
pre-synaptic neuron $k$ over time and their correspondent $S_k$ generate when
different values of time delay are considered. In Fig. \ref{fig2}(a), we plot
the membrane potential for the neuron $k$. Figures \ref{fig2}(b), \ref{fig2}(c),
and \ref{fig2}(d) exhibits the values of $S_k$ for $\tau=0$ ms, $\tau=6$ ms, and
$\tau=13$ ms, respectively. For a delay equal to zero (Fig. \ref{fig2}(b)), the
peak of $S_k$ match with the time in which the neuron $k$ spikes. For delay
greater than zero (Figs. \ref{fig2}(c) and (d)), the $S_k$ curve is shifted to
right by the value of the time delay $\tau$. Such effect is very significant for
the system dynamics, once it indicates that the synaptic current received by the
post-synaptic neuron $i$ is not instantaneous. It is important to mention that
time delays are expected for the type of connections considered in this work,
given by chemical synapses. Differently from the electric synapses, in which the
interactions among neurons are practically instantaneous due to the direct
transfer of ions, chemical synapses depend on the release of neurotransmitters,
which are associated with some delay until the post-synaptic neuron $i$ receives
the signal sent from the pre-synaptic neuron $k$ \cite{northrop01}. 

\begin{table}[htbp]
\caption{Description of the parameters, values, and units used in our numerical
simulations.} 
\centering
\begin{tabular}{lcc}
\hline
\small
\footnotesize \bf{Description} & \footnotesize \bf{Parameter} &\footnotesize
\bf{Values} \\
\hline
\footnotesize Number of neurons & $N$ & \footnotesize $100$ \\
\footnotesize Connection probability & $p$ & \footnotesize $0.1$\\
\footnotesize Membrane capacity & $C$ & \footnotesize 1 $\mu$F/cm$^2$ \\
\footnotesize Max. potassium conductance & $g_{\rm K}$ &  \footnotesize 36
mS/cm$^2$\\
\footnotesize Max. sodium conductance & $g_{\rm Na}$ & \footnotesize 120
mS/cm$^2$ \\
\footnotesize Max. leak conductance & $g_{\rm l}$ & \footnotesize 0.3 mS/cm$^2$
\\
\footnotesize Potassium reversal potential  & $V_{\rm K}$ & \footnotesize -77 mV
\\
\footnotesize Sodium reversal potential  & $V_{\rm Na}$ & \footnotesize 50 mV \\
\footnotesize Leak reversal potential & $V_{\rm l}$ & \footnotesize -54.4 mV \\
\footnotesize Exc. reversal potential & $V_{\rm r}^{\rm exc}$&  \footnotesize 20
mV \\
\footnotesize Exc. synaptic conductance& \footnotesize $g_{\rm exc}$ &
\footnotesize [0,1] mS/cm$^2$ \\
\footnotesize Const. ext. current density & $I^{\rm 0}_i$ & \footnotesize [10,14]
$\mu$A/cm$^2$ \\
\footnotesize Period that pulse is ON& \footnotesize $\Delta t^{\rm (ON)}$ &
[0,14] ms \\
\footnotesize Period that pulse is OFF& \footnotesize $\Delta t^{\rm (OFF)}$ &
[0,14] ms \\
\footnotesize Adjacency matrix & \footnotesize $A_{ik}$ & \footnotesize 0 or 1
\\
\footnotesize Time delay & $\tau$ & \footnotesize [0,14] ms \\
\footnotesize Decay time constant & $\tau_{\rm s}$ & \footnotesize 2.728 ms \\
\footnotesize Time step integration &  $\delta t$ & \footnotesize $10^{-2}$ ms\\
\footnotesize Initial time for analyses & $t_{\rm ini}$ & \footnotesize 5 s \\
\footnotesize Final time for analyses & $t_{\rm fin}$ & \footnotesize 10 s \\
\hline
\end{tabular}
\label{tabela}
\end{table}

Inspired by some works in the area that present a reasonable configuration to
mimic the behaviour of a neuronal network
\cite{borges17,hansen22,lameu21,borges16}, we build a network with $N=100$ HH
neurons randomly coupled with a probability of connections $p=0.1$. The initial
conditions are randomly distributed in $V_i\in[-80,0]$ mV and
$n_{i}=m_{i}=h_{i}=0$. A range of initial conditions is considered in order to
allow that the HH neurons to start their dynamics at different points from each
other, i.e., conditions in which some of these neurons spike more easily than
others. We also assume that $S_{k}$ is equal to $0$ for $t<t_k+\tau$. In our
simulations, the integration of the differential equations is done using the
fourth-order Runge-Kutta algorithm with a fixed integration time step
$\delta t=10^{-2}$ ms \cite{andreev19,borges16,butcher96}. A short summary about
the parameters, values, ranges, and units related to the neuronal description
of this work can be found along of Table \ref{tabela}
\cite{popovych14,lameu21,borges16}. 


\section{Diagnostics}
\label{sec3}

We consider a time interval from $t_{\rm ini}=5$ s to $t_{\rm fin}=10$ s, where
$t<t_{\rm ini}$ is the transient time. In our simulation, this time interval is
sufficient to perform analysis on the neuronal networks, since the system 
already presents stabilisation in the measurements. In order to extract an
average behaviour, we compute the means over a set of $100$ different numerical
simulations.

\subsection{Synchronisation}
\label{sec3.1}

The diagnostic method chosen to evaluate the level of synchronicity is the mean
value of the Kuramoto order parameter \cite{acebron05}, which is calculated as
\begin{equation}
\langle R\rangle=\frac{1}{t_{\rm fin}-t_{\rm ini}}\int_{t_{\rm ini}}^{t_{\rm fin}}\left|
\frac{1}{N} \sum_{i=1}^{N}{\rm exp} \left[{\rm j}\Phi_i(t)
\right]\right| dt,
\label{kuramoto}
\end{equation}
where ${\rm j}$ is an imaginary number defined as ${\rm j}=\sqrt{-1}$. The phase
$\Phi_i(t)$ is calculated by means of
\begin{equation}
\Phi_i(t)=2\pi m+2\pi\frac{t-t_{i}^{m}}{t_{i}^{m+1}-t_{i}^{m}}, 
\end{equation}
where $t_{i}^{m}$ is the time in which occurs the $m$-th spike of the neuron $i$.
The mean order parameter $\langle R \rangle$ is given in a range from $0$ to
$1$, where the synchronisation is identified when $\langle R \rangle \approx 1$.

\subsection{Mean synaptic current density}
\label{sec3.2}

The synaptic current density plays an important role in the connection among 
the HH neurons. With this in mind, we calculate the mean synaptic current
density for each time $t$ (after a transient) as 
\begin{equation}
\langle I^{\rm syn}(t) \rangle = \frac{1}{N} \sum_{i=1}^{N} I^{\rm syn}_{i}(t-\tau),
\end{equation}
where $I^{\rm syn}_i(t-\tau)$ is given by Eq. (\ref{I_syn}), while the mean value
for this time interval is given by
\begin{equation}
\langle I^{\rm syn} \rangle = \frac{1}{(t_{\rm fin}-t_{\rm ini})}
\int_{t_{\rm ini}}^{t_{\rm fin}} \langle I^{\rm syn}(t)\rangle dt.
\end{equation}

\subsection{Synaptic current distribution}
\label{sec3.3}

We define a measure $\zeta$ which indicates the distribution associated with the
shape of $\langle I^{\rm syn}(t)\rangle$. The $\zeta$ measure is written as
\begin{equation}
\zeta={{\rm mod}(H)\over{{\rm mean}(H)}},
\label{zeta}
\end{equation}
where mod$(H)$ and mean$(H)$ represent, respectively, the mode and the mean
value of the histogram $H$ associated with the shape of the time series of
$\langle I^{\rm syn}(t)\rangle$. In our simulations, $\zeta$ tends to $1$ if $H$
is gaussian-like, while $\zeta$ moves away from $1$ for asymmetric and non
gaussian-like $H$ histograms. The variation of $\zeta$ is related to the type of
the synchronisation level developed by the neuronal network.


\section{Time delay in unperturbed neuronal networks}
\label{sec4}

\begin{figure}[t]
\centerline{\includegraphics[width=0.5\linewidth]{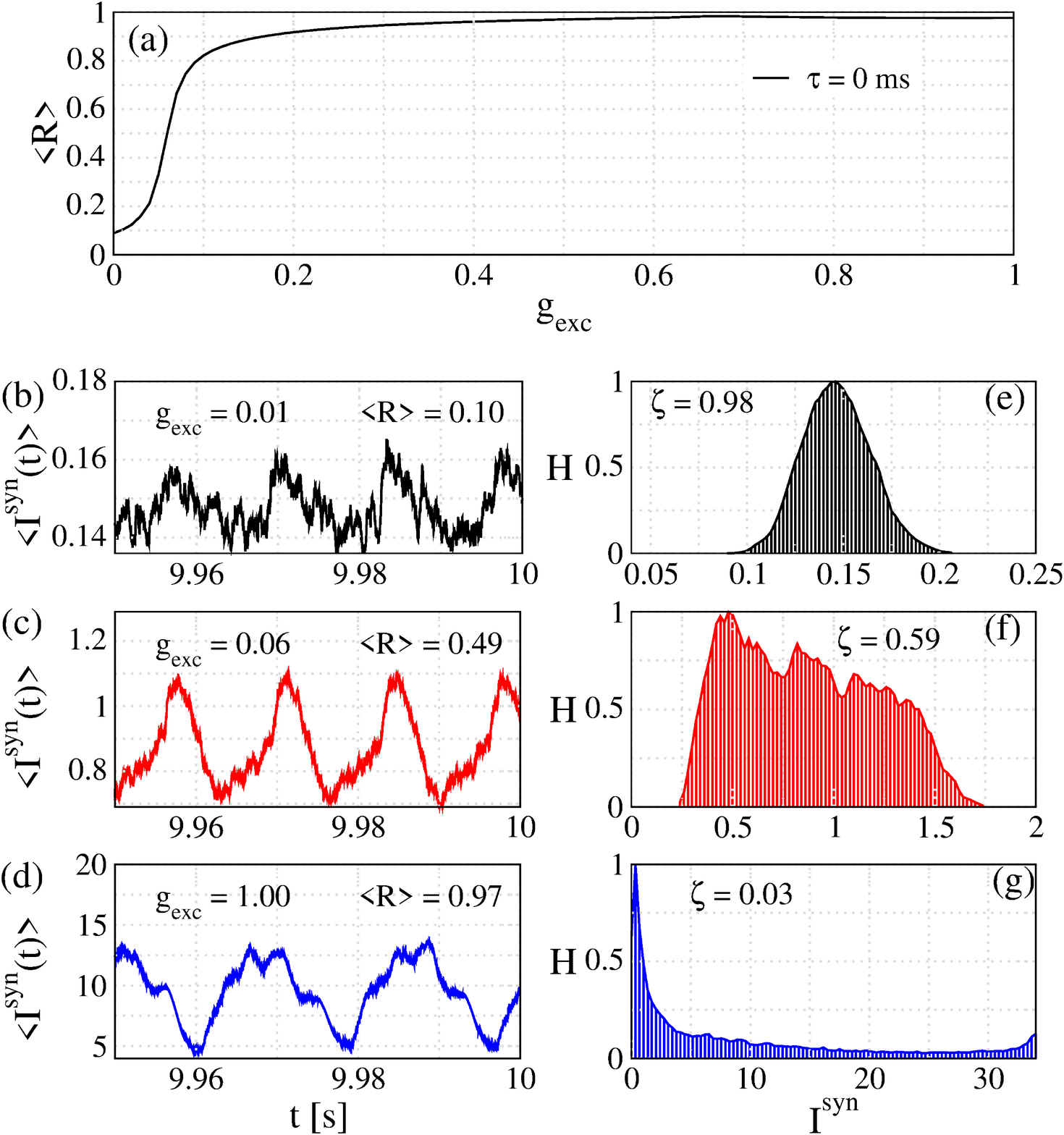}}
\caption{(a) Mean value of the Kuramoto order parameter as a function of
$g_{\rm exc}$. Time series of $\langle I^{\rm syn}(t) \rangle$ for (b)
$g_{\rm exc}=0.01$ mS/cm$^{2}$, (c) $g_{\rm exc}=0.06$ mS/cm$^{2}$, and (d)
$g_{\rm exc}=1.0$ mS/cm$^{2}$ with the respective histograms $H$ in the panels
(e), (f), and (g). We consider $100$ different numerical simulations for the HH
neurons and $\tau=0$ ms.}
\label{fig3}
\end{figure}

\begin{figure}[htbp]
\centerline{\includegraphics[width=0.5\linewidth]{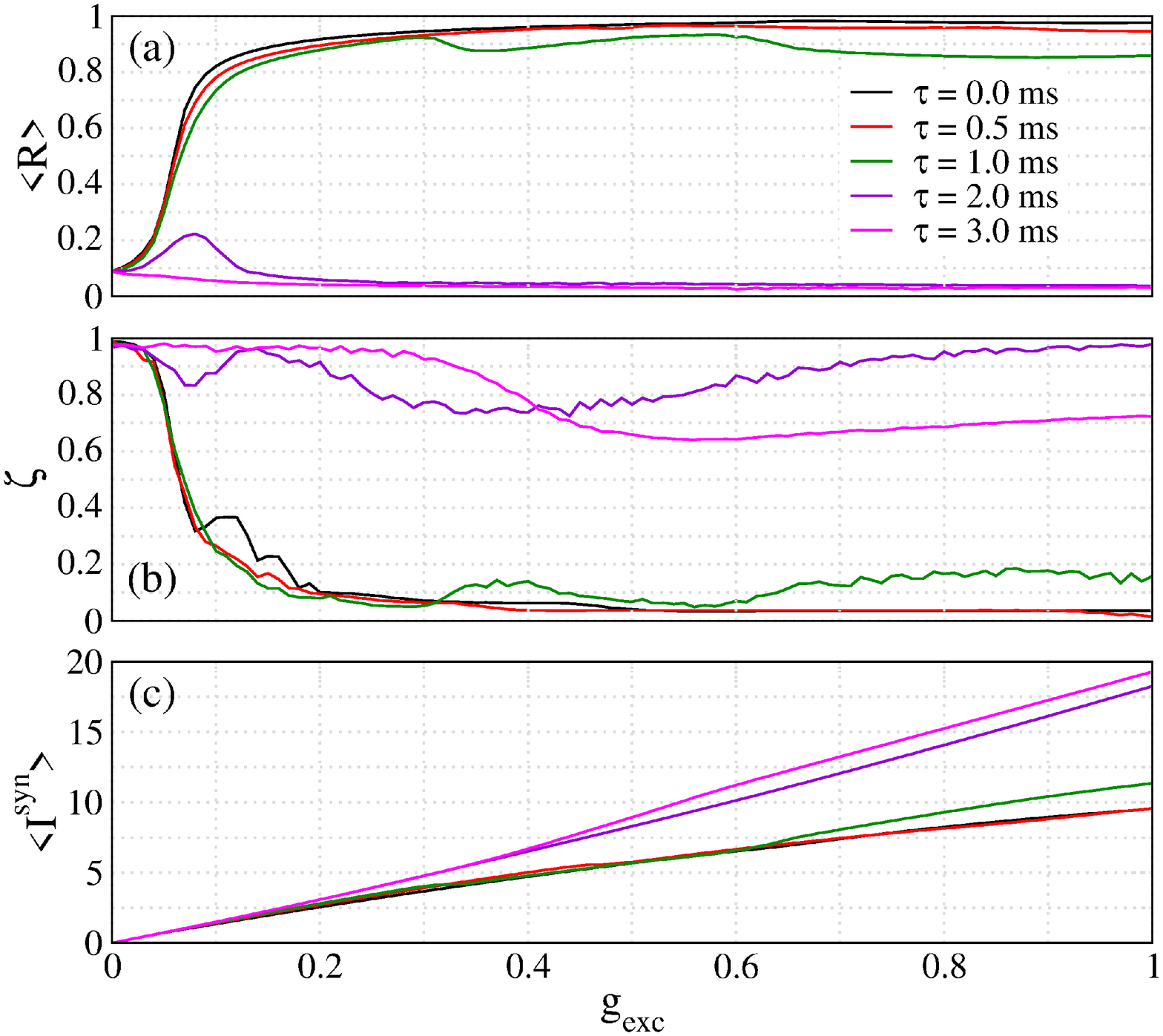}}
\caption{Three diagnostics as a function of the coupling strength $g_{\rm exc}$
for different valus of the time delay $\tau$. (a) Mean value of the Kuramoto
order parameter used in order to identify spike synchronisation of the HH
neurons. (b) $\zeta$ measure indicates the type of the histogram $H$ obtained
in the mean synaptic current $\langle I^{\rm syn}(t)\rangle$. (c) Mean synaptic
current obtained during the analyse ($\langle I^{\rm syn}\rangle$). We consider
$100$ different numerical simulations.}
\label{fig4}
\end{figure}

For $\tau=0$ ms and increasing the synaptic coupling $g_{\rm exc}$, we observe
that $\langle R\rangle$ increases, as shown in Fig. \ref{fig3}(a), where the
neurons go from desyncrhonised to synchronised activities. In Figs.
\ref{fig3}(b), \ref{fig3}(c), and \ref{fig3}(d), we plot the time series of
$\langle I^{\rm syn}(t)\rangle$ for (a) $g_{\rm exc}=0.01$ mS/cm$^2$, (b)
$g_{\rm exc}=0.06$ mS/cm$^2$), and (c) $g_{\rm exc}=1.0$ mS/cm$^2$. As one can see,
as greater is $g_{\rm exc}$, less noisy is the shape of
$\langle I^{\rm syn}(t)\rangle$. In order to extract a characteristic about these
different shapes, we compute a normalised (by the mode value) histogram $H$ of
$\langle I^{\rm syn} (t)\rangle$ obtained during a time series of the last $5$ s
of the numerical simulations. Figures \ref{fig3}(e), \ref{fig3}(f), and
\ref{fig3}(g) display the respective histograms $H$ associated with each
$\langle I^{\rm syn} (t)\rangle$. Comparing each one of these cases, it is
possible to see that when $\langle R\rangle$ is small (Fig. \ref{fig3}(b)), the
histogram is Gaussian-like (Fig. \ref{fig3}(e)). For spike synchronisation
(Fig. \ref{fig3}(d)), the histogram has a specific shape (Fig. \ref{fig3}(g)),
that is asymmetric and non Gaussian-like. For the intermediary 
case (Figure \ref{fig3}(f)), for instance $\langle R \rangle=0.49$, we see that
the histogram shape changes during the transition from desynchronous to
synchronous patterns. The diagnostic via $\zeta$ measure for the cases
discussed in Figs. \ref{fig3}(e), \ref{fig3}(f), and \ref{fig3}(g) are $0.98$,
$0.59$, and $0.03$, respectively. The Gaussian-like distribution shown in Fig.
\ref{fig3}(e) approaches to $1$, while another one go away from the unit value.
The complete comparison between the mean order parameter and the
characterisation from the spike desynchronisation to spike synchronisation via
$\zeta$ measure, when no time delay is considered, is displayed in Fig.
\ref{fig4}(a) and \ref{fig4}(b) (black line). The result indicates and confirms
that the mechanism involved behind the spike synchronisation, that is obtained
via the increasing of $g_{\rm exc}$, is linked with the alterations in the shape
of $\langle I^{\rm syn}(t)\rangle$. In addition, in Fig. \ref{fig4}(c), we observe
that $\langle I^{\rm syn}\rangle$ increases with $g_{\rm exc}$ (black line).

\begin{figure}[b]
\centerline{\includegraphics[width=0.5\linewidth]{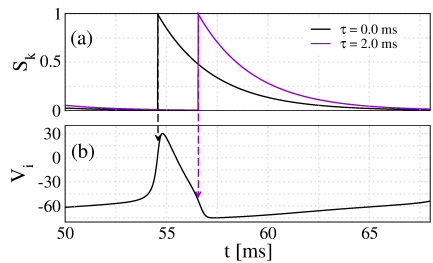}}
\caption{Schematic representation of different times in which the synaptic 
conductance $S_{k}$ from the pre-synaptic neuron $k$ achieves the pos-synaptic
neuron $i$.}
\label{fig5}
\end{figure}

\begin{figure*}[t]
\centerline{\includegraphics[width=0.8\linewidth]{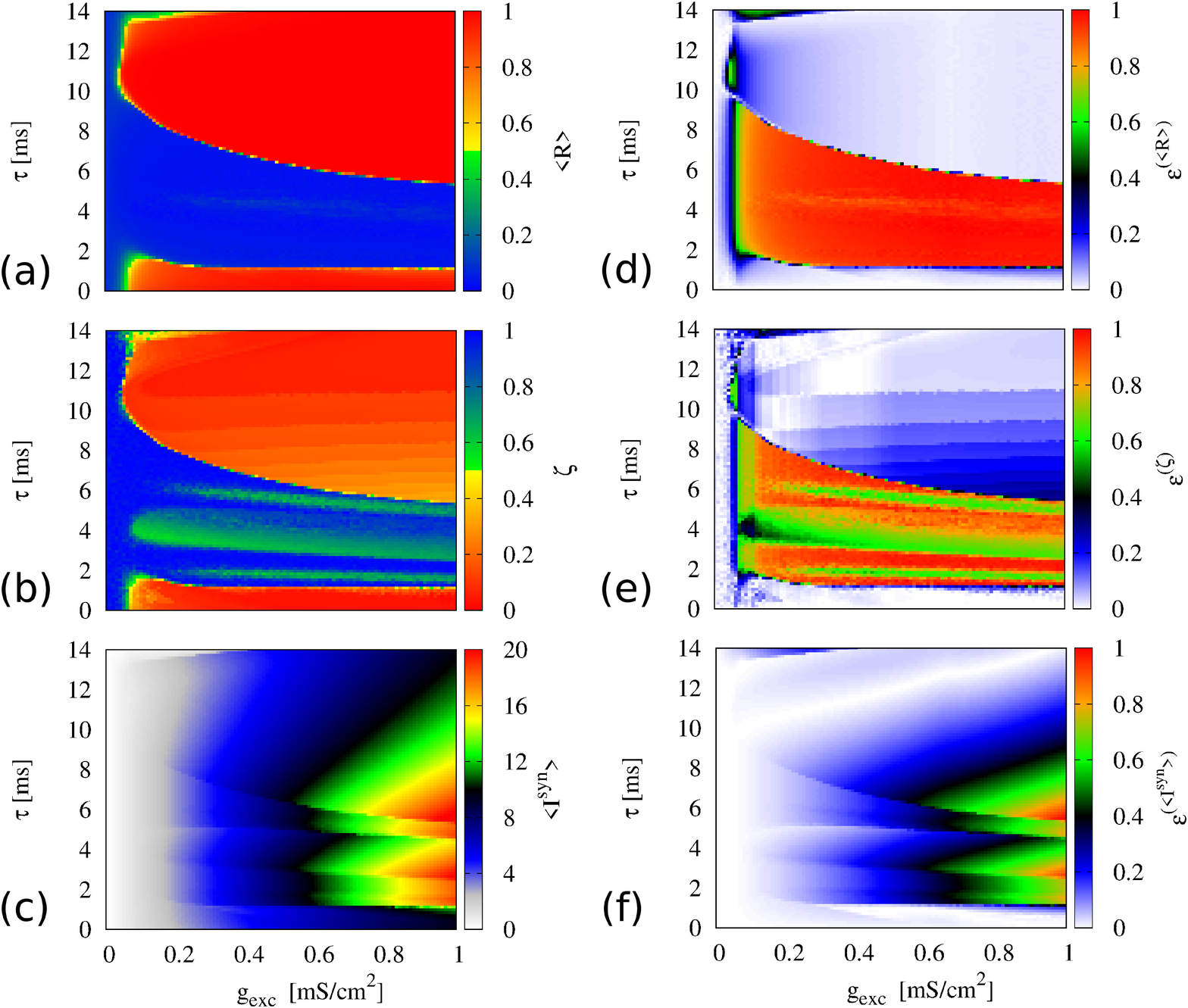}}
\caption{Parameter space $g_{\rm exc}\times\tau$ for (a) $\langle R\rangle$, (b)
the $\zeta$ measure related to the type of the histogram $H$ associated with
$\langle I^{\rm syn}(t)\rangle$, (c) $\langle I^{\rm syn}\rangle$, (d)
$\varepsilon^{(\langle R\rangle)}$, (e) $\varepsilon^{(\zeta)}$,  and (f) 
$\varepsilon^{(\langle I^{\rm syn}\rangle)}$. Depending on $\tau$, the panels (a), (b),
and (c) shows that the time delay can suppress the spike synchronisation. In the
panels (d), (e), and (f), the parameter spaces indicate that there are some
values of the time delay wich produce similar effects than for $\tau=0.0$ ms.}
\label{fig6}
\end{figure*}

Now, we focus on the cases in which $\tau>0$, denoted by the colour lines in 
Fig. \ref{fig4}. In Fig. \ref{fig4}(a) for $\tau\leq 1$ ms, it is possible to
see a small reduction in the $\langle R\rangle$ value for larger $g_{\rm exc}$.
Although the time delay changes the instant that the synaptic current density
arrives on the neurons, this time shift is not enough to promote relevant
alteration in the collective behaviour of the HH neurons and consequently in the
synchronisation. Therefore, for small time delays, the neuronal dynamics remain
closer to the case without time delay. However, for $\tau\ge 2$ ms, the time
shift is able to produce significant changes in the neuronal network, mainly
over the mean synaptic current density, leading to a strong suppression of
synchronised activities for all $g_{\rm exc}$ values. Figure \ref{fig4}(b) shows
that $\tau\ge 2$ ms increases $\zeta$, while $\tau<1$ ms cause only small
changes. In Figs. \ref{fig4}(a) and \ref{fig4}(b), it is possible to verify that
the transition between the desynchronous and synchronous states are linked once
again to changes in the shape of $\langle I^{\rm syn}(t)\rangle$, that is
characterised by the $\zeta$ measure. The changes are effects of the
introduction of the synaptic time delay in the neuronal connections. In Fig.
\ref{fig4}(c), we identify a relation between the synaptic time delay and the
value of $\langle I^{\rm syn}\rangle$. Depending on the value of $\tau$ and 
$g_{\rm exc}$, the system can exhibit an increase of $\langle I^{\rm syn}\rangle$ 
(colour lines) when compared with the case without time delay (black line). 
 
As illustrated in Fig. \ref{fig5}(a), if no time delay is considered, $S_{k}=1$ 
achieves the post-synaptic neuron $i$ when the potential membrane is
$V_i\approx30$ mV (black dashed arrow), which makes
$I_{i}^{\rm syn}\propto-10\times g_{\rm exc}$ (see Eq. (\ref{I_syn})). On the other
hand, for $\tau=2$ ms, $S_{k}=1$ achieves the post-synaptic neuron $i$ when
$V_{i}\approx-50$ mV (violet dashed arrow), producing
$I_{i}^{\rm syn}\propto 70\times g_{\rm exc}$, indicating an increase of the
synaptic current for the neuron $i$ at that moment. If a similar effect occurs
for more HH neurons in the network, it is reasonable to expected that, in
average, $\langle I^{\rm syn}\rangle$ might have its value increased. 

In Figs. \ref{fig6}(a), \ref{fig6}(b), and \ref{fig6}(c), we compute the
parameter space $g_{\rm exc}\times\tau$, where the colour scales indicate
$\langle R\rangle$, $\zeta$, and $\langle I^{\rm syn}\rangle$, respectively. We
consider $g_{\rm exc}\in[0,1]$ mS/cm$^2$ and $\tau\in[0,14]$ ms. The range of
$\tau$ is chosen in order to create a link with the mean inter-spike interval (ISI) 
of the HH neurons in the network (about $14$ ms). As can be seen in Fig. \ref{fig6}(a), for $0<\tau\leq1$ ms,
the effects of the time delay are small in the synchronisation, which leads 
$\langle R\rangle$ to remain similar to the case in which $\tau=0$ ms, in a way
almost independently of the $g_{\rm exc}$. However, for $1<\tau\lesssim 5.5$ ms, 
we observe an interval of $\tau$ in which the neuronal synchronisation is 
suppressed for almost all $g_{\rm exc}$. If $\tau>5.5$ ms, it is also possible 
to verify the appearance of synchronisation. Our simulations show that there is 
a preferential range of the time delay in which the synchronisation can be
suppressed. In Fig. \ref{fig6}(b), the transitions between desynchronised to
synchronised spikes can be identified by means of the $\zeta$ measure, while in
Fig. \ref{fig6}(c), it is possible to see that, depending on $\tau$ and
$g_{\rm exc}$, $\langle I^{\rm syn}\rangle$ can increase. For $\tau\approx 14$ ms,
the parameter spaces exhibit an appearance very close to the case in which the
network has no time delay. A similarity can be verified in Figs. \ref{fig6}(d),
\ref{fig6}(e), and \ref{fig6}(f), where we compute the parameter space 
$g_{\rm exc}\times\tau$ in colour scale $\varepsilon^{(y)}$.  The parameter
$\varepsilon^{(y)}$ can be written as 
\begin{equation}
\varepsilon^{(y)}={y(\tau)-y(0)\over{\varepsilon^{y}_{\rm Max}}},
\quad y=\langle R\rangle,\zeta, \langle I^{\rm syn}\rangle
\label{epsilon}
\end{equation}
where $y(\tau)$ represents the measure of $y$ for $\tau>0$, $y(0)$ is the
measure of $y$ for $\tau=0$, and $\varepsilon^{y}_{\rm Max}$ corresponds to the
maximum difference for this parameter with and without time delay in all
considered parameter space. For instance, if we consider $y=\langle R\rangle$,
$\varepsilon^{(\langle R\rangle)}\in [0,1]$ is the difference of $\langle R\rangle$
between a network with $\tau=0$ ms and $\tau>0$ ms. If
$\varepsilon^{(\langle R\rangle)}\approx0$, the network dynamics almost does not
suffer changes due to the time delay. However, if
$\varepsilon^{(\langle R\rangle)}\approx 1$, the opposite is observed. As indicated
by Figs. \ref{fig6}(d), \ref{fig6}(e), and \ref{fig6}(f) for $\tau>0$ and
$\tau\leq 1$ ms, the neuronal dynamics has a high similarity with the result for
$\tau$ null. For the case in which $1<\tau<5.5$ ms, there are relevant
alterations in the neuron dynamics due to the effect of the time delay.
Considering $\tau>5.5$ ms, it is possible to observe that some parts of the
parameter space exhibit similarities with the case in which no time delay is
considered. Dynamically, this effect can be explained by Figs. \ref{fig2}(b),
\ref{fig2}(c), and \ref{fig2}(d). As the time delay increases, the $S_{k}$
function associated with the post-synaptic neuron $i$ exhibits a shift in
$\tau$ milliseconds. As largers is $\tau$, further it is the $S_{k}$ from the
original point ($\tau=0$ ms), indicated by the green dashed line in Fig.
\ref{fig2}. However, for $\tau=13$ ms, the $S_{k}$ is very far from the original
point (green dashed line), but it is delivered to the post-synaptic neuron $i$
almost at the same time in which a new spike of the pre-synaptic neuron $k$
occurs (orange dashed line). There is a kind of resonance due to the synaptic
current density, indicating that even delayed, the synapse produces an
equivalent effect of a neuronal network with instantaneous synaptic current
density, namely a network with no time delay. Our results suggest that the
values of $\tau$, which are able to produce the suppression of synchronised
activities, are in the interval $1<\tau<7$ ms, being more or less effective
depending on $g_{\rm exc}$. 


\section{Time delay effect in perturbed neuronal networks}
\label{sec5}

Recently, a great interest in the effect of external perturbations in neuronal
networks is getting the attention of the scientific community. External
perturbations can not only induce spike synchronisation, but also reduce the
abnormal synchronous behaviour \cite{paulo19,jesus19}. Through experimental analyses in
rats, Cota et al. \cite{cota21} in 2021 reported that nonperiodic electrical
stimulations can be a promising alternative for the treatment of epilepsy
crises. Chatterjee and Robert \cite{Robert} demonstrated that if some amount of 
noise is introduced into a stimulus, it is possible to improve the auditory
perception in cochlear implants. With this in mind, we perform numerical
analysis for a delayed network when three different types of pulsed
perturbations (periodical, random and mixed) are introduced \cite{hansen22}. We
study how the ranges of time delay can affect neuronal synchronous behaviour.

\subsection{Periodic pulses}
\label{sec4.1}

\begin{figure}[htbp]
\centerline{\includegraphics[width=0.5\linewidth]{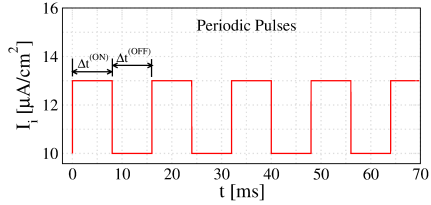}}
\caption{Schematic representation of periodic pulses applied over $I_{i}^{0}=10$
$\mu$A/cm$^2$, considering a pulse amplitude $\Gamma=3$ $\mu$A/cm$^2$ and
$\Delta t^{\rm (ON)}=\Delta t^{\rm (OFF)}=8$ ms.}
\label{fig7}
\end{figure}

We begin our analysis for the case in which the pulsed perturbation is
periodically applied over the neurons over time. In our simulations, we consider
a perturbation $\xi(t)$ in the external current density, in order that $I_{i}$
is given as 
\begin{equation}
I_{i}=I_{i}^{0}+\xi(t),
\label{external1}
\end{equation}
where $\xi(t)$ represents the term which assumes an amplitude equal to $0$ or
$\Gamma$, in an on-off configuration over time, generating a pulse profile. In
the periodic pulse, the time in which the pulse is on ($\Delta t^{\rm (ON)}$) and
off ($\Delta t^{\rm (OFF)}$) is the same. Figure \ref{fig7} displays a schematic
representation of a periodic pulsed perturbation with $I_{i}^{0}=10$
$\mu$A/cm$^2$, $\Gamma=3$ $\mu$A/cm$^2$ and
$\Delta t^{\rm (ON)}=\Delta t^{\rm (OFF)}=8$ ms.

In order to study the effect of pulsed perturbation over the neuronal network, 
we define two scenarios: (i) a network weakly coupled ($g_{\rm exc}=0.05$
mS/cm$^2$) and (ii) a network strongly coupled ($g_{\rm exc}=1.0$ mS/cm$^2$). An
intermediary scenario ($g_{\rm exc}=0.5$ mS/cm$^2$) is also investigated, however,
the numerical results found can be, without loss of generality, approached to
the case (ii). In both scenarios, we apply the pulsed perturbation with a
$\Gamma$ amplitude, where the associated time interval assumes
$\Delta t^{\rm (ON)}=\Delta t^{\rm (OFF)}=\Delta t\in[0,14]$ ms. The pulses assume
the on-off configurations in scales around the mean inter-spike interval (ISI)
of the HH neurons in the network.

\begin{figure*}[t]
\centerline{\includegraphics[width=0.8\linewidth]{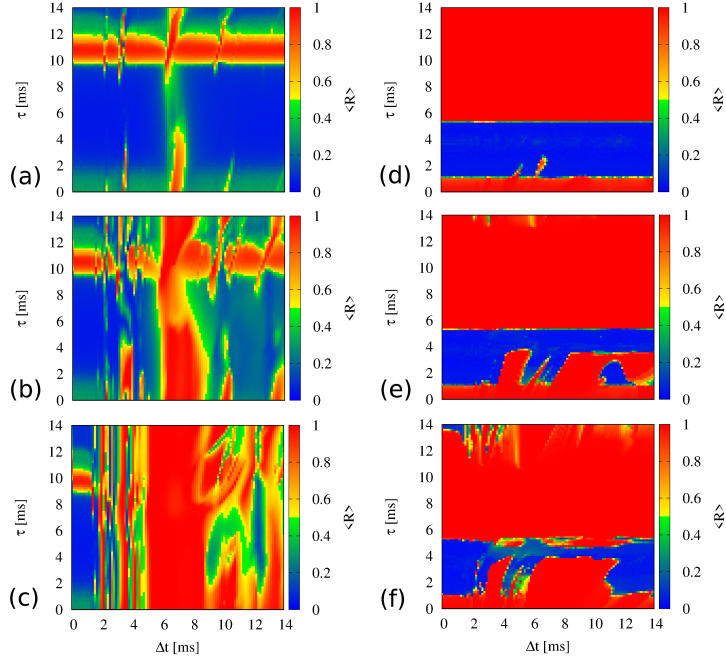}}
\caption{Parameter space $\Delta t\times\tau$ with the mean value of the
Kuramoto order parameter $\langle R\rangle$ in colour scale. In the panels (a),
(b), and (c), we show an analyse for a scenario of weak coupling
($g_{\rm exc}=0.05$ mS/cm$^{2}$) in which the HH neurons are perturbed by periodic
pulses with amplitude $\Gamma=1$ $\mu$A/cm$^2$, $\Gamma=3$ $\mu$A/cm$^2$, and
$\Gamma=10$ $\mu$A/cm$^2$, respectively. In the panels (d), (e), and (f), we
consider the same pulse amplitudes, respectively, however, considering a 
scenario of strong coupling ($g_{\rm exc}=1.0$ mS/cm$^{2}$). Depending on
$\Delta t$, we observe that the periodic perturbation is able to reduce the
ranges in wich $\tau$ has high effectiveness on the suppression of synchronised
activities, even when a small amplitude of the pulse is considered. For
$\Delta t\sim7$ ms, the interval of each cycle (on-off) of the pulse
perturbation, approximately, coincides with the mean inter-spike interval (ISI),
i.e., $2\Delta t$ ($14$ ms)=$\Delta t^{\rm(ON)}$ ($7$ ms) + $\Delta t^{\rm(OFF)}$
($7$ ms) $\approx$ ISI ($14$ ms), creating a kind of resonance that might be
inducing spike synchronisation.}
\label{fig8}
\end{figure*}

Figures \ref{fig8}(a), \ref{fig8}(b), and \ref{fig8}(c) display
$\langle R\rangle$ in the parameter space $\Delta t\times\tau$ for a weakly
coupled, where we consider $\Gamma=1$ $\mu$A/cm$^2$, $\Gamma=3$ $\mu$A/cm$^2$,
and $\Gamma=10$ $\mu$A/cm$^2$, respectively. If $\Delta t$ is approximately
lower than $2$ ms, the periodic pulses are not able to alter the dynamic of the
network, independently of the $\Gamma$ value. However, when $\Delta t>2$ ms, 
it is possible to observe spike synchronisation due to the pulsed perturbations.
Increasing the pulse amplitude, the range around $\Delta t\sim 7$ ms begins to
be more relevant to induce spike synchronisation. In this case, the interval of
each cycle (on-off) of the pulse perturbation, approximately, coincides 
with the mean inter-spike interval (ISI), i.e., 
$2\Delta t$ ($14$ms)=$\Delta t^{\rm(ON)}$ ($7$ ms) + $\Delta t^{\rm(OFF)}$ ($7$ ms)
$\approx$ ISI ($14$ ms). This fact is clear in Fig. \ref{fig8}(a), where small
amplitudes of pulsed perturbation are sufficient to induce synchronisation for
$\Delta t\sim7$ ms. In Fig. \ref{fig8}(e), where $\Gamma=10$ $\mu$A/cm$^2$, we
find ranges approximately in $5<\Delta t<8$ ms in which the pulse is able to
induce all the HH neurons in spike synchronisation, for all $\tau$ considered in
this study. 

Figures \ref{fig8}(d), \ref{fig8}(e), and \ref{fig8}(f) displays
$\langle R\rangle$ in $\Delta t\times\tau$ for strong coupling and the same
$\Gamma$ amplitudes considered in the weak coupling. In Figs. \ref{fig8}(d),
we observe that $\Gamma=1$ $\mu$A/cm$^2$ induces only small regions of the spike
synchronisation in the parameter space for $\Delta t$ around $7$ ms. Therefore,
under strong coupling, the neuronal network exhibits a greater resistance or a
lesser influence of small external perturbations. On the other hand, as the
amplitude $\Gamma$ increases, the perturbation starts to be more capable to
induce spike synchronisation, as shown in Figs. \ref{fig8}(e) and \ref{fig8}(f).
As can be seen in Fig. \ref{fig8}(f) for $\Gamma=10$ $\mu$A/cm$^2$, there is a
large continuous range of $6<\Delta t<10$ ms which reduces the capacity to
observe desynchronised spikes for some delays.

The periodic pulses can reduce the ranges of the time delay which are able to
suppres spike synchronisation in weakly and strongly coupled neurons. However,
such reduction depends on the time intervals in that the pulses are applied. Our
results suggest that for appropriate time intervals $\Delta t^{\rm (ON)}$ and 
$\Delta t^{\rm (OFF)}$, the transition from desynchronised activities to
synchronised ones can be done by pulses with high or low amplitude, indicating
that these parameters of the pulse have a crucial role in the alterations of the
collective neuronal behaviour.

\subsection{Random pulses}
\label{sec4.2}

\begin{figure}[htpb]
\centerline{\includegraphics[width=0.5\linewidth]{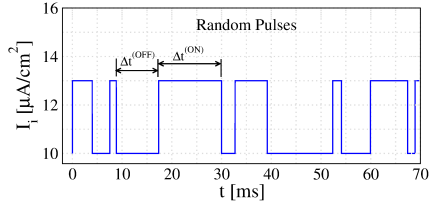}}
\caption{Schematic representation of the random pulse dynamics applied over
$I_{i}^{0}=10$ $\mu$A/cm$^2$, considering $\Gamma=3$ $\mu$A/cm$^2$ with
$\Delta t^{\rm (ON)}$ and $\Delta t^{\rm (OFF)}$ randomly chosen (following a
uniform distribution) in $[0;14]$ ms.}
\label{fig9}
\end{figure}

We consider a pulsed perturbation according to a random protocol for the choice
of the time in which the pulses are on-off. We define that $\Delta t^{\rm (ON)}$
and $\Delta t^{\rm (OFF)}$ are randomly chosen (following a uniform distribution)
in $[0,14]$ ms. Figure \ref{fig9} exhibits a schematic representation of the
random pulses over time for $I_{i}^{0}=10$ $\mu$A/cm$^2$ and $\Gamma=3$
$\mu$A/cm$^2$, where $\Delta t^{\rm (ON)}$ and $\Delta t^{\rm (OFF)}$ assume random
values.

\begin{figure}[t]
\centerline{\includegraphics[width=0.5\linewidth]{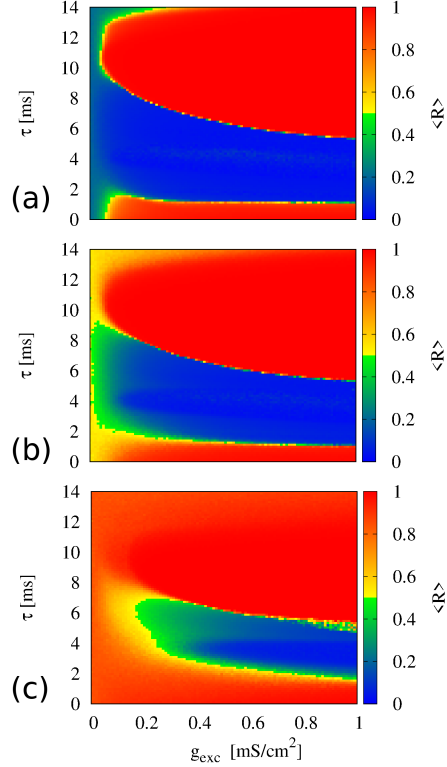}}
\caption{Parameter space $g_{\rm exc}\times\tau$ with the mean value of the
Kuramoto order parameter $\langle R\rangle$ in colour scale. In the panels
(a), (b), and (c), we plot an analyse for the parameter space when random
pulses with amplitude $\Gamma=1$ $\mu$A/cm$^2$, $\Gamma=3$ $\mu$A/cm$^2$, and
$\Gamma=10$ $\mu$A/cm$^2$ are applied over the HH neurons, respectively. We see
that the random perturbation is able to reduce the ranges in which $\tau$ has
high effectiveness on the suppression of synchronised activities. However,
differently of the periodic case, the random pulses need to assume $\Gamma$ with
larger amplitudes in order to promote changes in the delayed network.}
\label{fig10}
\end{figure}

In order to study the effects of random pulses in the neuronal dynamics, we
compute the parameter space. Figure \ref{fig10}(a), \ref{fig10}(b), and
\ref{fig10}(c) display $\langle R\rangle$ in the parameter space
$g_{\rm exc}\times\tau$ for $\Gamma=1$ $\mu$A/cm$^2$, $\Gamma=3$ $\mu$A/cm$^2$,
and $\Gamma=10$ $\mu$A/cm$^2$, respectively. For $\Gamma=1$ $\mu$A/cm$^2$, the
neuronal network has no significant alterations, independently of $g_{\rm exc}$ 
or $\tau$ used in this work. If the pulse amplitude is increased to 
$\Gamma=3$ $\mu$A/cm$^2$, the network begins to exhibit some changes, however,
only for small values of $g_{\rm exc}$. In this case, the random pulses can
improve the level of spike synchronisation in the delayed neuronal network
approximately for $\tau<2$ ms and $\tau>10$ ms. On the other hand, as
$g_{\rm exc}$ increases, the synchronisation does not show any remarkable changes,
indicating a difference between the periodic and random pulses. For the periodic
case and appropriate time intervals (Figs. \ref{fig8}(a), \ref{fig8}(b),
\ref{fig8}(d), and \ref{fig8}(e)), the pulses with $\Gamma=1$ $\mu$A/cm$^2$ and
$\Gamma=3$ $\mu$A/cm$^2$ are enough to promote alterations in the networks with
weak and strong couplings. In Fig. \ref{fig10}(c), if $\Gamma=10$ $\mu$A/cm$^2$,
there are more synchronised ranges in the parameter space, including in strong
couplings, what is not verified for random pulses with small amplitudes
(Figs. \ref{fig10}(a) and \ref{fig10}(b)).

Our results show that random pulses can also reduce ranges of $\tau$ associated
with desynchronised activities in the parameter space. However, such reduction
is related to the amplitude values of the pulses. For the random pulses, 
$\Gamma$ needs to assume large values in order to induce synchronised regions 
along the parameter space. For the periodic pulses, even small amplitudes of
$\Gamma$, if applied in appropriate time intervals, are enough to change the
neuronal dynamics.

\subsection{Mixed pulses}
\label{sec4.3}

\begin{figure}[htbp]
\centerline{\includegraphics[width=0.5\linewidth]{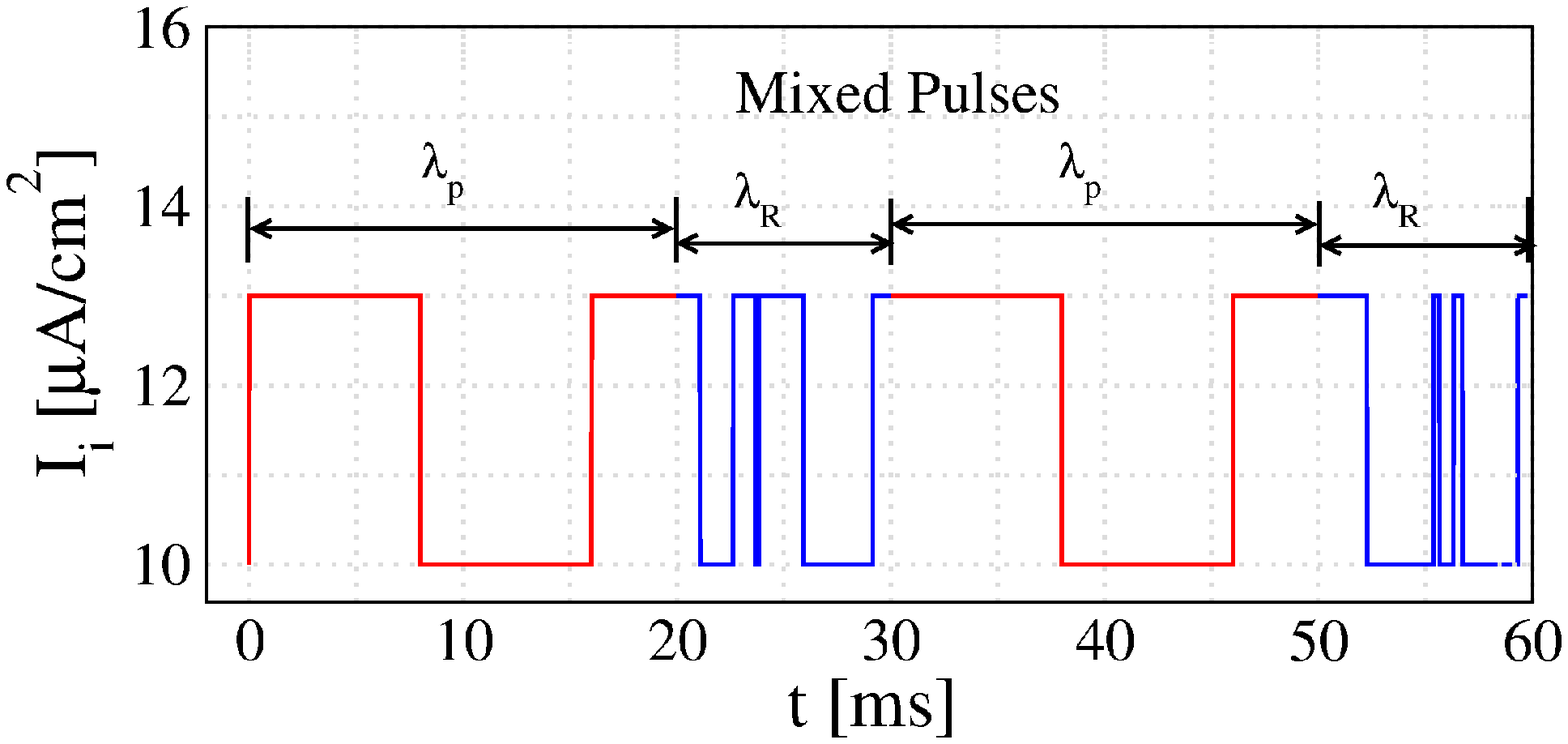}}
\caption{Schematic representation of the mixed pulse for $I_{i}^{0}=10$
$\mu$A/cm$^2$, taken into account a pulse amplitude $\Gamma=3$ $\mu$A/cm$^2$.
We consider a $\lambda_{\rm P}=20$ ms with
$\Delta t^{\rm (ON)}=\Delta t^{\rm (OFF)}=8$ ms for the periodic part of the pulse.
In the random part, we consider $\lambda_{\rm R}=10$ ms with $\Delta t^{\rm (ON)}$
and $\Delta t^{\rm (OFF)}$ randomly chosen in $[0,14]$ ms. }
\label{fig11}
\end{figure}

We consider that the mixed perturbation is composed of sequential time windows 
$\lambda_{\rm P}$ and $\lambda_{\rm R}$, where the pulses are assumed as periodic 
and random, respectively. In our numerical simulations, we define the relation
about the sizes of these time windows as
\begin{equation}
\lambda_{\rm P}=\lambda_{\rm P}^{0}-\lambda_{\rm R},
\label{windows}
\end{equation}
where $\lambda_{\rm P}^0$ and $\lambda_{\rm R}$ are the sizes of the time window
for the periodic and random pulses, respectively. For $\lambda_{\rm R}=0$ ms, the
pulse is formed only by sequences of windows $\lambda_{\rm P}=\lambda^0_{\rm P}$,
configuring a complete periodic pulse. On the other hand, if we consider
$\lambda_{\rm R}>0$ ms, the perturbation is composed of alternating sequence
windows, $\lambda_{\rm P}$ and $\lambda_{\rm R}$, where the pulses assume the
periodical and random profiles, respectively. Consequently, if
$\lambda_{\rm R}=\lambda_{\rm P}^{0}$, only random pulses are observed, once there
are no time windows $\lambda_{\rm P}$ in the signal. In Fig. \ref{fig11}, we
show a sketch of this mixed pulses for $I_{i}^{0}=10$ $\mu$A/cm$^2$, $\Gamma=3$
$\mu$A/cm$^2$, $\Delta t^{\rm (ON)}=\Delta t^{\rm (OFF)}=8$ ms (for the periodic part 
of the pulse), $\lambda_{\rm P}=20$ ms, and $\lambda_{\rm R}=10$ ms.

\begin{figure}[htbp]
\centerline{\includegraphics[width=0.5\linewidth]{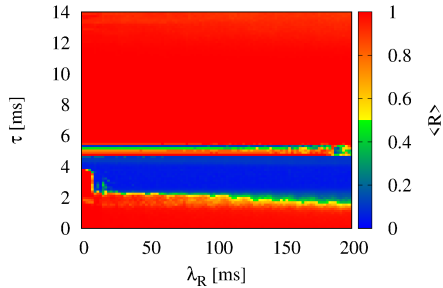}}
\caption{Parameter space $\lambda_{\rm R}\times\tau$  with the mean value of the 
Kuramoto order parameter $\langle R\rangle$ in colour scale. We see the
transition from a complete periodic pulsed perturbation ($\lambda_{\rm R}=0$ ms) 
to a fully random pulsed perturbation ($\lambda_{\rm R}=200)$ ms). For
$\lambda_{\rm R}\approx 5$ ms, the mixed pulse begins to induce effects
characteristic of random pulses.}
\label{fig12}
\end{figure}

In order to analyse the transition from a complete periodic pulse to a random
one, we consider $\Delta t^{\rm (ON)}=\Delta t^{\rm (OFF)}$ $=8$ ms, $\Gamma=10$
$\mu$A/cm$^2$, and $\lambda_{\rm P}^{0}=200$ ms. In Fig. \ref{fig12}, we calculate
the parameter space $\lambda_{\rm R}\times\tau$ with $\langle R\rangle$ in colour
scale. When the network is perturbed only by periodic pulses
($\lambda_{\rm R}=0$ ms), the region of desynchronised spike activity is reduced.
For complete random pulses ($\lambda_{\rm R}=200$ ms), there are more values of
$\tau$ that are able to suppress spike synchronisation. We find that 
$\lambda_{\rm R}\approx 5$ ms is enough to change the parameter space in order to
increase the interval of $\tau$ in which occurs suppression of spike
synchronisation. Increasing $\lambda_{\rm R}$, the range of $\tau$ becomes
larger. Moreover, as indicated by our numerical simulation for this case, the
transition between a complete periodic pulse to a roandom one is given in a
smooth way. 

Our results indicate that small windows of random pulses embedded in a sequence
of periodic ones can be enough to influence the neuronal dynamics more like 
a random perturbation than a periodic one. The neurons behave more similarly to
the case in which the network is under random perturbations. This analyse is
interesting and complement a recent observation which indicates that same 
effects appear in a scenario of weak coupling \cite{hansen22}.


\section{Conclusions}
\label{sec6}

In this work, we study the effects of time delays in the synaptic conductance
as a way to suppress spike synchronisation developed in coupled Hodgkin Huxley
neurons. The time delays in the synaptic conductance are related to no
instantaneous transmission of the synaptic currents between the neurons.
Depending on the time delay and coupling strength values, changes in the
synaptic current can induce or not spike synchronisation. Our results show that
there is an important range of time delays ($\tau>1$ until $\tau\approx 5.5$ ms)
in which the synchronised activities are suppressed, independently of the
coupling strength value. 

We analyse how a delayed neuronal network behaves when pulsed perturbations are
applied in the neurons. Our results indicate that both periodic and random
pulses can reduce the intervals of the time delay values in which the spike
synchronisation is suppressed. For appropriate time intervals, the periodic
pulses are able to generate synchronisation in the presence 
of time delays associated with desynchronized activities, even for pulses with
small amplitudes. In this case, we find a type of resonance related to the
intervals of pulsed cycles close to the mean inter-spike interval
(ISI) of the HH neurons. For the random case, alterations in the synchronisation
can be only observed for larger amplitude of the pulses. Our simulations
demonstrate that if small windows of random pulses are embedded in a sequence of
periodic pulses, the mixed perturbations can exhibit similar characteristics to
the random pulses. 

Considering that spike synchronisation can be associated with some brain
pathologies, such as epilepsy, then the search for alternative methods that
aims to avoid synchronous pattern are needed. As suggested by our findings in
this work, the time delay can be an approach to reduce or even avoid spike
synchronisation in generic neuronal networks, especially if the time delay is
from $\tau>1$ ms to $\tau\approx 5.5$ ms. In addition, such an interval is also
capable of holding desynchronised spikes even for some perturbation conditions 
applied over the HH neurons, which is an important result due to the fact that
perturbations can be associated for instance with some sensory stimulus. In
such a scenario, the delay might be part of important steps toward to find an
useful and applicable method to control synchronised activities, mostly
considering that time delay is a comum effect in chemical synapses. In this way,
interventions that can change the time of neuronal transmission could bring new
possibilities for synchronisation control.


\section*{Acknowledgements}

The authors acknowledge the financial support from S\~ao Paulo Research
Foundation (FAPESP, Brazil) (Grants Nos. 2015/50122-0, 2018/03211-6,
2019/ 09150-1, 2020/ 04624-2), National Council for Scientific and Technological
Development (CNPq), the Coordena\c c\~ao de Aperfei\c coa\-mento de Pessoal de
N\'ivel Superior - Brasil (CAPES) and Funda\c c\~ao Arauc\'aria.



\end{document}